\documentclass[twocolumn,amsmath,amssymb,pre]{revtex4}

\usepackage{graphicx}
\usepackage{bm}
\usepackage{amssymb}
\usepackage{amsmath}
\usepackage{amsfonts}

\usepackage[utf8]{inputenc}

\begin{document}

\title{Three-dimensional solitons in fractional nonlinear Schr\"{o}dinger equation
 with exponential saturating nonlinearity}

\author{Volodymyr M. Lashkin}
\email{vlashkin62@gmail.com} \affiliation{$^1$Institute for
Nuclear Research, Pr. Nauki 47, Kyiv 03028, Ukraine}
\affiliation{$^2$Space Research Institute, Pr. Glushkova 40 k.4/1,
Kyiv 03187,  Ukraine}

\author{Oleg K. Cheremnykh}
\affiliation{Space Research Institute, Pr. Glushkova 40 k.4/1,
Kyiv 03187, Ukraine}

%\date{\today}

\begin{abstract}
We study the fractional three-dimensional (3D) nonlinear
Schr\"{o}dinger equation with exponential saturating nonlinearity.
In the case of the L\'{e}vy index $\alpha=1.9$, this equation can
be considered as a model equation to describe strong Langmuir
plasma turbulence. The modulation instability of a plane wave is
studied, the regions of instability depending on the L\'{e}vy
index, and the corresponding instability growth rates are
determined. Numerical solutions in the form of 3D fundamental
soliton (ground state) are obtained for different values of the
L\'{e}vy index. It was shown that in a certain range of soliton
parameters it is stable even in the presence of a sufficiently
strong initial random disturbance, and the self-cleaning of the
soliton from such initial noise was demonstrated.

\medskip
\medskip

\end{abstract}

\maketitle

\section{Introduction}

The concept of a fractional derivative $\partial^{\alpha}/\partial
x^{\alpha}$ is a generalization of the ordinary derivative
$\partial^{n}/\partial x^{n}$ to the case of a real number
$\alpha$, and has a rather long history (more than three hundred
years) \cite{Oldham1974,Samko1993,Miller1993,Podlubny1999,Das2011}
that began back in 1695 in the correspondence of Leibnitz with
L'Hopital, where the case of $\alpha=1/2$ was considered. Physical
applications of the fractional derivative are very diverse and
include such areas and phenomena as fractional quantum mechanics,
fractional and strange kinetics, fluid mechanics, optics,
electromagnetics, diffusion-reaction processes, anomalous
transport, fractals, and a number of others
\cite{Agrawal2004,Tarasov2010,Herrmann2011,Uchaikin2013,Malomed2021,
Malomed2024,Mihalache2024,Kevrekidis2024}.
The concept of fractional quantum mechanics, which stimulated many
works on physical applications of the fractional derivative, was
formulated by Laskin in \cite{Laskin2000,Laskin2002,Laskin2018},
and the fractional quantum mechanical Schr\"{o}dinger equation was
introduced by generalizing the Feynman path integral over L\'{e}vy
trajectories corresponding random L\'{e}vy flights in the theory
of Brownian motion. In this case, the number $\alpha$ (L\'{e}vy
index) characterizing the fractionality of the Laplacian in the
Schr\"{o}dinger equation takes values in the range $0<\alpha\leq
2$.

The fractional Laplacian (with the Riesz fractional derivative) is
defined as
\begin{gather}
(-\Delta)^{\alpha/2}f(\mathbf{x})=\mathcal{F}^{-1}(|\mathbf{k}|^{\alpha}\mathcal{F}f)
\nonumber \\
 =\frac{1}{(2\pi)^{D}}\int
|\mathbf{k}|^{\alpha}\hat{f}(\mathbf{k})\mathrm{e}^{i\mathbf{k}\cdot\mathbf{x}}d^{D}\mathbf{k},
\label{Laplas1}
\end{gather}
where $\hat{f}=\mathcal{F}f$ is the Fourier transform of $f$, and
$D$ is the spatial dimension ($D=1,2,3$). Another definition of
the fractional Laplacian (with the Caputo fractional derivative)
often used in applications is
\begin{gather}
(-\Delta)^{\alpha/2}f(\mathbf{x})=\frac{2^{\alpha-1}\alpha \Gamma
((D+\alpha)/2)}{\sqrt{\pi^{D}}\Gamma (1-\alpha/2)}
\nonumber \\
\times \, \mathrm{P}.\,\mathrm{V}.\int
\frac{f(\mathbf{x})-f(\mathbf{x}^{\prime})}{|\mathbf{x}-\mathbf{x}^{\prime}|^{D+\alpha}}
d^{D}\mathbf{x}^{\prime},   \quad \mathrm{for} \quad 0<\alpha\leq
2,
 \label{Laplas2}
\end{gather}
where P. V. stands for the Cauchy principal value. Both of these
definitions are equivalent in Schwartz space
\cite{Samko1993,Kwasnicki2017}. Note that there is also a fairly
large number of other definitions of the fractional derivative,
both classical and new ones introduced quite recently
\cite{Oliveira2014}. In this paper, for the fractional Laplacian
we use Eq. (\ref{Laplas1}).

One of the most physically significant models with fractional
derivatives is the fractional nonlinear Schr\"{o}dinger equation
(NLS). It generalizes the well-known NLS equation to the case of a
fractional Laplacian. Physical applications of the fractional NLS
equation in most cases concern nonlinear optics (see, e.g.,
reviews \cite{Malomed2021,Malomed2024} and references therein).
The fractional Davey-Stewartson equations (one of the
generalizations of the two-dimensional NLS equation) with the
L\'{e}vy index $\alpha=3/4$ was obtained in \cite{Davey2015} for
nonlinear surface water waves.

Numerical solutions of the one-dimensional (1D) fractional NLS
equation with cubic nonlinearity in the form of a fundamental
soliton (ground state) for different L\'{e}vy indices $0<\alpha<2$
were found in \cite{Klein2014}. There, the evolution of an initial
profile different from the soliton one was also studied, and both
focusing and defocusing nonlinearity were considered. In the
focusing case, at $\alpha=1$ there is a collapse in the so-called
critical regime (negative Hamiltonian and the 1D norm exceeds the
soliton 1D norm), and for $0<\alpha<1$ in the supercritical regime
(a sufficiently large 1D norm). The 1D fractional NLS equation
with cubic nonlinearity was also studied in detail in
\cite{Chen2018}. In particular, analytical soliton solutions were
obtained using a variational approach for different values of
$\alpha$. It is shown that solitons are stable for $1<\alpha<2$,
but for $\alpha=1$ the soliton collapses. Various problems within
the framework of the fractional NLS equation (modulation
instability, the presence of a trapping potential, two-dimensional
structures, including vortex solitons, etc.) were considered in
\cite{Malomed2021,Malomed2024,Zhang2017,Malomed2020,Ovolabi2016}.
The possible applicability of the inverse scattering transfrom
method for the 1D fractional NLS equation was discussed in
\cite{Ablowitz2022}.

In connection with the problem of Langmuir plasma turbulence
\cite{Zakharov1972,Thornhill1978,Goldman1984}, in
\cite{Adjemyan1989} the three-dimensional (3D) stochastic NLS
equation with external noise was studied using quantum field
theory approach. In particular, using the quantum field
renormalization group method, it was shown that in this case the
usual linear dispersion in the NLS equation $\omega=k^{2}$
(assuming $\psi\sim\exp (i\mathbf{k}\cdot\mathbf{x}-i\omega t)$)
is replaced by $\omega=k^{2-\gamma_{a}}$, where $\gamma_{a}$ is
the so-called anomalous dimension known from quantum field theory,
and $\gamma_{a}=0.0804$. The effective NLS equation with such a
corrected dispersion no longer contains a stochastic term. This
equation appears to be the only example of a fractional
three-dimensional NLS equation that has physical applications. As
is known, an arbitrary initial perturbation with a negative
Hamiltonian or formal stationary solutions of the conventional
three-dimensional NLS equation with cubic nonlinearity collapse
(blow-up), that is, become singular in a finite time
\cite{Sulem1999}. The same is true (as shown by the 1D examples
given above) for the fractional three-dimensional NLS equation
with $\alpha <2$. In reality, the formation of a singularity is
prevented by dissipation or higher-order nonlinearities. For
nonlinear Langmuir waves, the collapse can be arrested for the
exponential saturable nonlinearity corresponding to the Boltzmann
distribution of plasma particles \cite{Laedke1984,Lashkin2020}.

In this paper, we study the three-dimensional fractional NLS
equation with exponential saturating nonlinearity and numerically
obtain soliton solutions (ground states) for different values of
the L\'{e}vy index. We pay special attention to the case of the
Levy index $\alpha=2-\gamma_{a}\sim 1.9$ corresponding to an
important physical application in the problem of Langmuir plasma
turbulence. By direct numerical simulation, we show the stability
of solitons in a certain range of parameters satisfying the
Vakhitov-Kolokolov criterion. For the first time, the phenomenon
of self-cleaning of a stable soliton from a sufficiently strong
initial random disturbance is demonstrated for the fractional NLS
equation.

The paper is organized as follows. The basic model equation is
presented in Sec. II. In Sec. III, we consider modulation
instability of a plane wave and find the corresponding instability
thresholds and instability growth rates. Sec. IV, soliton
solutions are found, and a numerical analysis of the stability of
solitons is performed. Finally, Sec. V concludes the paper.

\section{Model equation}

We consider the fractional nonlinear Schr\"{o}dinger equation with
exponential saturable nonlinearity in spatial dimension $D=3$,
\begin{equation}
\label{main} i\frac{\partial \psi}{\partial
t}-(-\Delta)^{\alpha/2}\psi+\left[1-\exp
(-|\psi|^{2})\right]\psi=0.
\end{equation}
As noted in the Introduction, the fractional Laplacian with
$\alpha=2-\gamma_{a}\sim 1.9$ in Eq. (\ref{main}) corresponds to
the corrected linear dispersion law (in fact, a renormalized
propagator) of Langmuir waves for the stochastic NLS equation with
cubic nonlinearity in the theory of Langmuir plasma wave
turbulence. On the other hand, the exponential nonlinearity in the
three-dimensional NLS equation with the usual Laplacian
($\alpha=2$) prevents the phenomenon of wave collapse and can
results in the existence of stable nonlinear coherent structures
such as solitons and vortex solitons (they can be treated as
"elementary bricks" of strong Langmuir turbulence). In particular,
Eq. (\ref{main}) with $\alpha=1.9$ can be considered as a model
equation for Langmuir turbulence without the phenomenon of
catastrophic collapse.

Equation (\ref{main}) conserves the 3D norm
\begin{equation}
\label{Energy} N=\int |\psi|^{2}d^{3}\mathbf{x},
\end{equation}
and Hamiltonian
\begin{equation}
\label{Hamiltonian} H=\!\!\int\!\left\{
\mathrm{Re}\!\left[\psi^{\ast}(-\Delta)^{\alpha/2}\psi\right]-|\psi|^{2}-\exp
(-|\psi|^{2})+1\right\}d^{3}\mathbf{x},
\end{equation}
and can be written in the hamiltonian form
\begin{equation}
i\frac{\partial\psi}{\partial t}=\frac{\delta
H}{\delta\psi^{\ast}}.
\end{equation}

The evolution of an arbitrary initial disturbance within the
framework of Eq. (\ref{main}) occurs under the influence of two
competing factors - dispersion and nonlinearity. Dispersion causes
the wave packet to spread and, in this sense, counteracts
collapse. However, as $\alpha$ decreases, the dispersion becomes
increasingly unable to arrest the collapse (blow-up). For example,
at $\alpha=1$, even one-dimensional solitons of the fractional NLS
equation with cubic nonlinearity are unstable and collapse. At
sufficiently large amplitudes $|\psi|\gg 1$, the nonlinear term in
Eq. (\ref{main}) effectively becomes linear. This is an example of
the so-called saturable nonlinearity (not necessarily
exponential), which is often encountered, for example, in
nonlinear optics \cite{Kivshar_book2003}. Saturable nonlinearities
cause collapse arrest (or beam self-focusing in two-dimensional
models) and can lead to the existence of stable coherent
structures.

\section{Modulational instability}

Equation (\ref{main}) has an exact solution in the form of a
monochromatic plane wave
\begin{equation}
\label{plane1} \psi=|\psi_{0}|\exp
(i\mathbf{k}_{0}\cdot\mathbf{x}-i\omega_{0} t)
\end{equation}
with a frequency depending on the amplitude $\psi_{0}$,
\begin{equation}
\label{plane2} \omega_{0}=k_{0}^{\alpha}+\exp (-|\psi_{0}|^{2})-1.
\end{equation}
Consider the stability of such a plane wave. The perturbed plane
wave solution has the form
\begin{equation}
\label{pert1} \psi=(|\psi_{0}|+\delta\psi)\exp
(i\mathbf{k}_{0}\cdot\mathbf{x}-i\omega_{0} t),
\end{equation}
where
\begin{equation}
\label{pert2}
\delta\psi=\psi^{+}\mathrm{e}^{i\mathbf{k}\cdot\mathbf{x}-i\Omega
t}+\psi^{-}\mathrm{e}^{-i\mathbf{k}\cdot\mathbf{x}+i\Omega t},
\end{equation}
is a linear modulation with the frequency $\Omega$ and the wave
vector $\mathbf{k}$. Linearizing Eq. (\ref{main}) in $\delta\psi$,
we get the nonlinear dispersion relation
\begin{gather}
1-\left[1-\exp (-|\psi_{0}|^{2})\right]
\left[\frac{1}{\omega_{\mathbf{k}_{0}
+\mathbf{k}}-\omega_{\mathbf{k}_{0}}-\Omega}\right. \nonumber \\
\left.
 +\frac{1}
{\omega_{\mathbf{k}_{0}-\mathbf{k}}-\omega_{\mathbf{k}_{0}}+\Omega}\right]=0,
\label{nonlin-disp}
\end{gather}
where $\omega_{\mathbf{q}}=|\mathbf{q}|^{\alpha}$ is the linear
dispersion relation for Eq. (\ref{main}). In the case of
short-wave modulations $\mathbf{k}\gg\mathbf{k}_{0}$, and taking
into account that $\omega_{\mathbf{k}}$ is even function, Eq.
(\ref{nonlin-disp}) becomes
\begin{equation}
\label{nonlin-disp-pure}
\Omega^{2}=\omega_{\mathbf{k}}\left\{\omega_{\mathbf{k}}-2[1-\exp
(-|\psi_{0}|^{2})]\right\}.
\end{equation}
Equation (\ref{nonlin-disp-pure}) predicts a purely growing
instability (modulational instability) provided
\begin{equation}
2[1-\exp (-|\psi_{0}|^{2})]>k^{\alpha},
\end{equation}
that is, when the amplitude threshold is exceeded,
\begin{equation}
\label{region} |\psi_{0}|>\sqrt{-\ln (1-k^{\alpha}/2)}.
\end{equation}
The instability growth rate $\gamma$ is given by
\begin{equation}
\label{pure} \gamma=\sqrt{2k^{\alpha}[1-\exp
(-|\psi_{0}|^{2})]-k^{2\alpha}}.
\end{equation}
\begin{figure}
\includegraphics[width=3.4in]{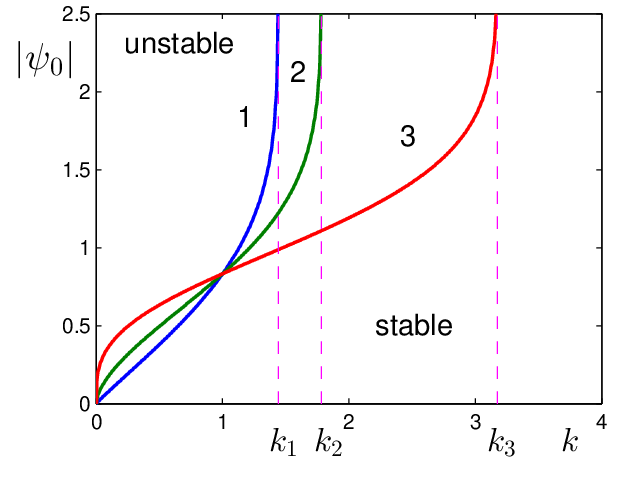}
\caption{\label{fig1} Modulational instability region on the plane
$(|\psi_{0}|,k)$. The areas above the corresponding curves
correspond to instability. The numbers above the curves correspond
to different $\alpha$: 1 - $\alpha=1.9$, 2 - $\alpha=1.2$, and 3 -
$\alpha=0.6$. The corresponding critical wave numbers
$k_{1}=1.44$, $k_{2}=1.78$, and $k_{3}=3.17$ are indicated.}
\end{figure}
\begin{figure}
\includegraphics[width=3.4in]{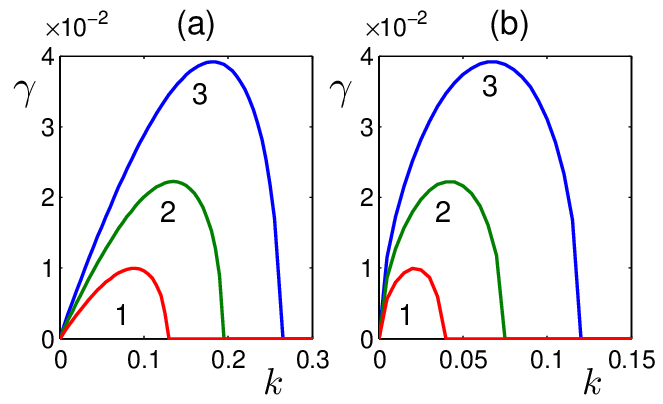}
\caption{\label{fig2} Dependence of the instability growth rate
$\gamma$ on wave number $k$ for different amplitude values
$\psi_{0}$; (a) $\alpha=1.9$; (b) $\alpha=1.2$. The numbers under
the curves correspond to different amplitudes: 1 -
$|\psi_{0}|=0.1$, 2 - $|\psi_{0}|=0.15$, and 3 -
$|\psi_{0}|=0.2$.}
\end{figure}
For each fixed value of $\alpha$, instability occurs only for wave
numbers $k$ exceeding the critical value
$k_{\mathrm{cr}}=2^{1/\alpha}$, and at $k\rightarrow
k_{\mathrm{cr}}$ the instability threshold tends to infinity. The
region of instability determined by Eq. (\ref{region}) on the
plane $(k,|\psi_{0}|)$ is shown in Fig.~\ref{fig1} for different
values of $\alpha$, namely $\alpha=1.9$, $\alpha=1.2$ and
$\alpha=0.6$. The corresponding values of the critical wave number
$k_{\mathrm{cr}}$ are equal to $k_{1}=1.44$, $k_{2}=1.78$ and
$k_{3}=3.17$, respectively. The optimal wave number of
perturbations, that is, corresponding to the maximum instability
growth rate in Eq. (\ref{pure}) is $k_{\mathrm{opt}}=[1-\exp
(-|\psi_{0}|^{2})]^{1/\alpha}$, and the corresponding instability
growth rate is
\begin{equation}
\label{max-growth} \gamma_{m}=1-\exp (-2|\psi_{0}|^{2}),
\end{equation}
and does not depend on $\alpha$. Dependence of the instability
growth rate $\gamma$ on wave number $k$ for different amplitude
values $\psi_{0}$ in the cases $\alpha=1.9$ and $\alpha=1.2$ is
shown in Fig.~\ref{fig2}.

In the opposite case of short-wave modulations with
$\mathbf{k}\ll\mathbf{k}_{0}$, using
\begin{equation}
\omega_{\mathbf{k}_{0}\pm\mathbf{k}}\sim
\omega_{\mathbf{k}_{0}}\pm \frac{\partial
\omega_{\mathbf{k}_{0}}}{\partial
\mathbf{k}_{0}}\cdot\mathbf{k}+\frac{1}{2}\frac{\partial^{2}
\omega_{\mathbf{k}_{0}}}{\partial \mathbf{k}_{0}^{2}}k^{2},
\end{equation}
from Eq. (\ref{nonlin-disp}) we obtain
\begin{equation}
\label{nonlin-disp-convective} (\Omega-
\mathbf{k}\cdot\mathbf{v}_{g})^{2}=\frac{1}{4}(\omega_{\mathbf{k}_{0}}^{\prime\prime})^{2}k^{4}
-\omega_{\mathbf{k}_{0}}^{\prime\prime}k^{2}[1-\exp
(-|\psi_{0}|^{2})],
\end{equation}
where
\begin{equation}
\mathbf{v}_{g}=\frac{\partial \omega_{\mathbf{k}_{0}}}{\partial
\mathbf{k}_{0}}=\frac{\partial (\mathbf{k}_{0}\cdot
\mathbf{k}_{0})^{\alpha/2}}{\partial\mathbf{k}_{0}}=\alpha
k_{0}^{\alpha-2}\mathbf{k}_{0}
\end{equation}
is the group velocity, and
$\omega_{\mathbf{k}_{0}}^{\prime\prime}=\alpha
(\alpha-1)k_{0}^{\alpha-2}$. If the amplitude threshold
$|\psi_{0}|$ determined by
\begin{equation}
4[1-\exp
(-|\psi_{0}|^{2})]>\omega_{\mathbf{k}_{0}}^{\prime\prime}k^{2},
\end{equation}
is exceeded, then Eq. (\ref{nonlin-disp-convective}) corresponds
to convective instability when growing disturbances are carried
away with the group velocity $\mathbf{v}_{g}$, and the instability
growth rate is given by
\begin{equation}
\gamma=2k\sqrt{\omega_{\mathbf{k}_{0}}^{\prime\prime}[1-\exp
(-|\psi_{0}|^{2})-\omega_{\mathbf{k}_{0}}^{\prime\prime}k^{2}/4]}.
\end{equation}

\section{Three-dimensional soliton }

We look for stationary solutions of Eq. (\ref{main}) of the form
\begin{equation}
\label{stationary} \psi (\mathbf{r},t)=u(\mathbf{r})\exp (i\lambda
t),
\end{equation}
where $\lambda>0$ is a free parameter, and the function
$u(\mathbf{r})$ is assumed to be real without loss of generality.
Then from Eq. (\ref{main}) we have
\begin{equation}
\label{main1} -\lambda u-(-\Delta)^{\alpha/2}u+\left[1-\exp
(-u^{2})\right]u=0.
\end{equation}
An analytical solution of Eq. (\ref{main1}) is impossible, but in
addition, due to the exponential form of nonlinearity and the
impossibility of analytical calculation of the corresponding
integrals, the variational approach used for the fractional NLS
equation with cubic nonlinearity
\cite{Malomed2021,Chen2018,Qiu2020}, is also apparently
inapplicable. Note that the solutions of the fractional NLS
equation ($\alpha\neq 2$) with general nonlinearities decay
algebraically at infinity as $\sim 1/|\mathbf{x}|^{D+\alpha}$
\cite{Frank2013,Frank2016,Li2024}, in contrast to the case
$\alpha=2$, where the solutions are localized exponentially.
\begin{figure}
\includegraphics[width=3in]{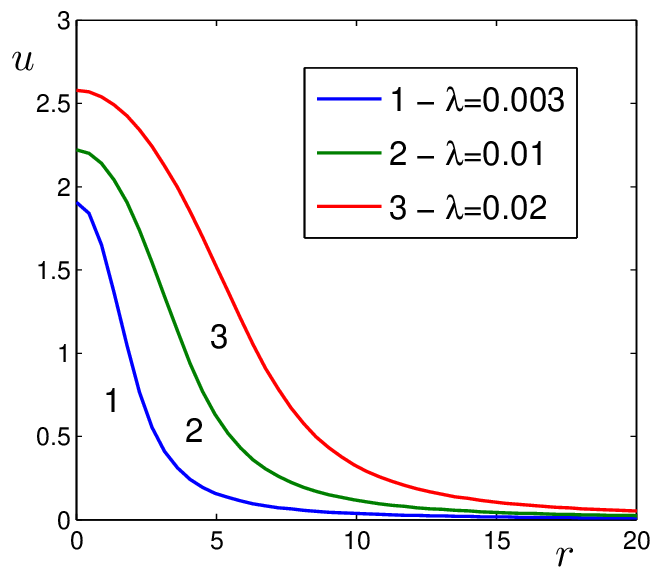}
\caption{\label{fig3} Radial profiles of a spherically symmetric
soliton (ground state) for the case of the L\'{e}vi index
$\alpha=1.2$ for different values of $\lambda$.}
\end{figure}
\begin{figure}
\includegraphics[width=3in]{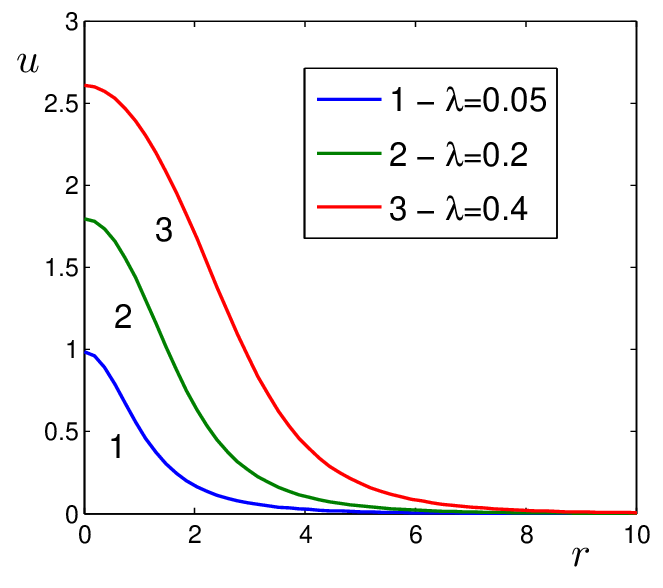}
\caption{\label{fig4} Radial profiles of a spherically symmetric
soliton (ground state) for the case of the L\'{e}vi index
$\alpha=1.9$ for different values of $\lambda$.}
\end{figure}

\begin{figure}
\includegraphics[width=3.4in]{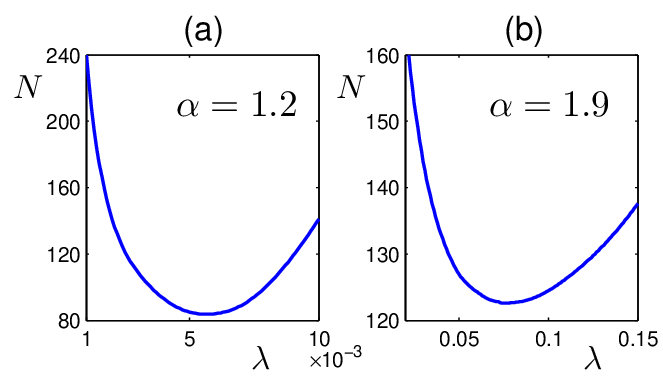}
\caption{\label{fig5} The 3D norm $N$ of the fundamental soliton
as a function of the parameter $\lambda$: (a) L\'{e}vy index
$\alpha=1.2$; (b) L\'{e}vy index $\alpha=1.9$.}
\end{figure}

We numerically find localized solutions of Eq. (\ref{main1}) using
the Petviashvili method
\cite{Petviashvili1976,Petviashvili_book1992,Lakoba2007}. Periodic
boundary conditions are assumed and, due to the algebraic behavior
at infinity, the box length is taken to be sufficiently large that
the solution at the boundary is negligible small. The advantage of
the Petviashvili method is that it is used in Fourier space (that
is, the fractional Laplacian is introduced in a simple natural
way) and, in combination with the Fast Fourier Transform (FFT),
does not require much computational time even on very high
resolution grids. Note that there is a modification of this method
\cite{Lashkin2008_77,Lashkin2008_78} that uses only physical space
and is therefore applicable to equations containing an explicit
dependence on spatial variables, but in this case the method is
slower (FFT is not used). After the Fourier transform, defined
here for an arbitrary function $f(\mathbf{r})$ as
\begin{equation}
f_{\mathbf{k}}=\int f(\mathbf{r})\exp
(-i\mathbf{k}\cdot\mathbf{r})d^{3}\mathbf{r},
\end{equation}
equation (\ref{main1}) is written in the form
\begin{equation}
G_{\mathbf{k}}^{-1}u_{\mathbf{k}}=N_{\mathbf{k}},
\end{equation}
where $G_{\mathbf{k}}=-1/(\lambda+k^{\alpha})$ and
$N_{\mathbf{k}}$ accounts for the nonlinear term. Then the
Petviashvili iteration procedure at the $n$-th iteration is
\begin{equation}
u_{\mathbf{k}}^{(n+1)}=sG_{\mathbf{k}}N_{\mathbf{k}}^{(n)},
\end{equation}
where $s$ is the so-called stabilizing factor determined by
\begin{equation}
s=\left(\frac{\int |u_{\mathbf{k}}^{(n)}|^{2}d^{3}\mathbf{k}}{\int
u_{\mathbf{k}}^{\ast,(n)}G_{\mathbf{k}}N_{\mathbf{k}}^{(n)}
d^{3}\mathbf{k}}\right)^{\gamma},
\end{equation}
and the parenthetic superscript denotes the iteration step index.
The nonlinear term at each step was first calculated in physical
space and then its Fourier transform $N_{\mathbf{k}}$ was used.
For power-law nonlinearity $u^{p}$, the fastest convergence occurs
for
\begin{equation}
\label{gamma} \gamma=\frac{p}{p-1}.
\end{equation}
Petviashvili formulated this in the form of a mnemonic rule, but
the choice of the optimal value of $\gamma$ was rigorously
justified in Ref.~\cite{Pelinovsky2004}. The procedure always
converges to the nonlinear ground state, i. e. fundamental
soliton, regardless of the initial guess. Moreover, the rate of
convergence is almost independent of the initial approximation.
For nonlinearity other than power-law, the value of $\gamma$
corresponding to the fastest convergence is chosen empirically and
$1<\gamma<p/(p-1)$, where $p$ is the smallest exponent in the
Taylor series expansion of nonlinearity. For the exponential
nonlinearity in Eq. (\ref{main1}), we chose $\gamma=1.3$. We used
$u (\mathbf{r})=\lambda^{1/2}\exp (-\lambda^{1/2}r^{2})$ as the
initial guess in all runs. The iterations rapidly converge to a
three-dimensional spherically symmetric soliton solution. The
progressive iterations were terminated when the value $|s-1|$ fell
below $10^{-15}$. Radial profiles of the three-dimensional soliton
solutions for L\'{e}vy index $\alpha=1.2$ and for L\'{e}vy index
$\alpha=1.9$ at different values of $\lambda$ are shown in
Fig.~\ref{fig3} and Fig.~\ref{fig4}, respectively.

\begin{figure}
\includegraphics[width=3.4in]{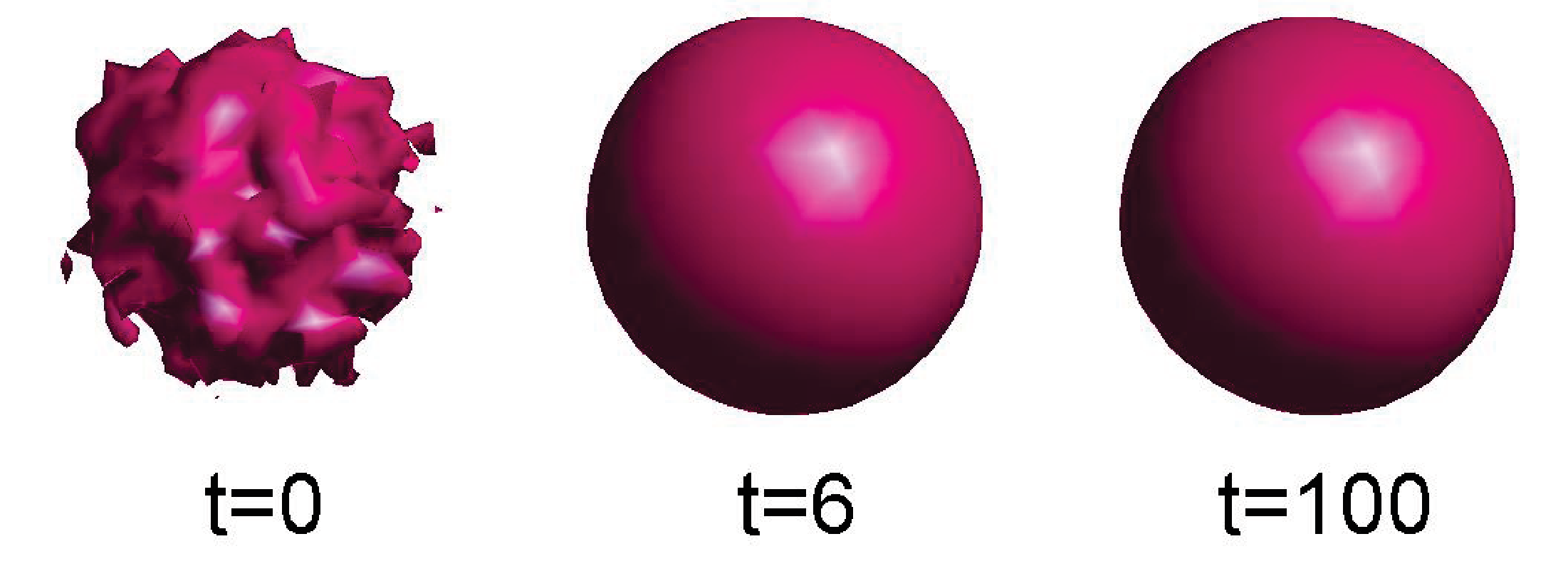}
\caption{\label{fig6} Self-cleaning and stable dynamics of the
fundamental soliton for the L\'{e}vy index $\alpha=1.9$ and
$\lambda=0.15$. The initial state is perturbed by the white noise
with the parameter $\varepsilon=0.05$; the shape (isosurface
$|\psi|=1.35$) of the perturbed soliton at the initial moment
$t=0$, at $t=6$, and at $t=100$. }
\end{figure}

\begin{figure}
\includegraphics[width=3.4in]{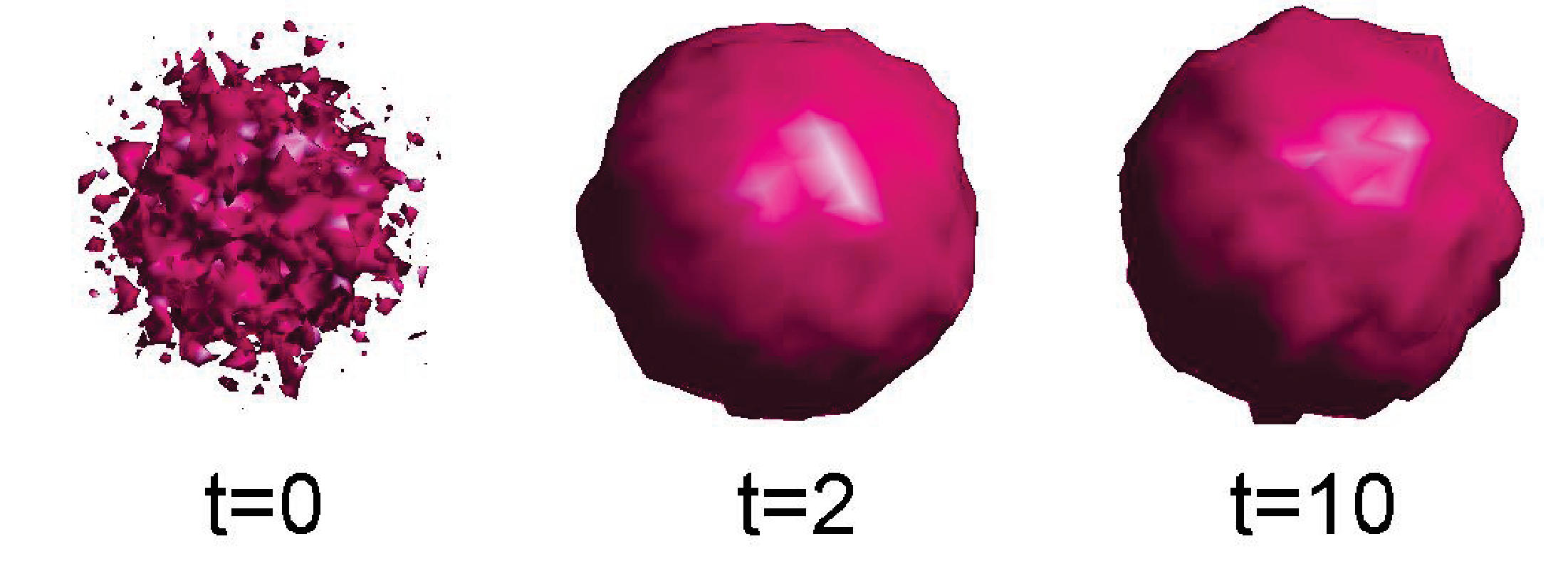}
\caption{\label{fig7} Evolution of the fundamental soliton for the
L\'{e}vy index $\alpha=1.9$ and $\lambda=0.15$ with a sufficiently
strong random initial perturbation ($\varepsilon=0.3$); the shape
(isosurface $|\psi|=1.35$) of the perturbed soliton at the initial
moment $t=0$, at $t=2$, and at $t=10$. }
\end{figure}

\begin{figure}
\includegraphics[width=3in]{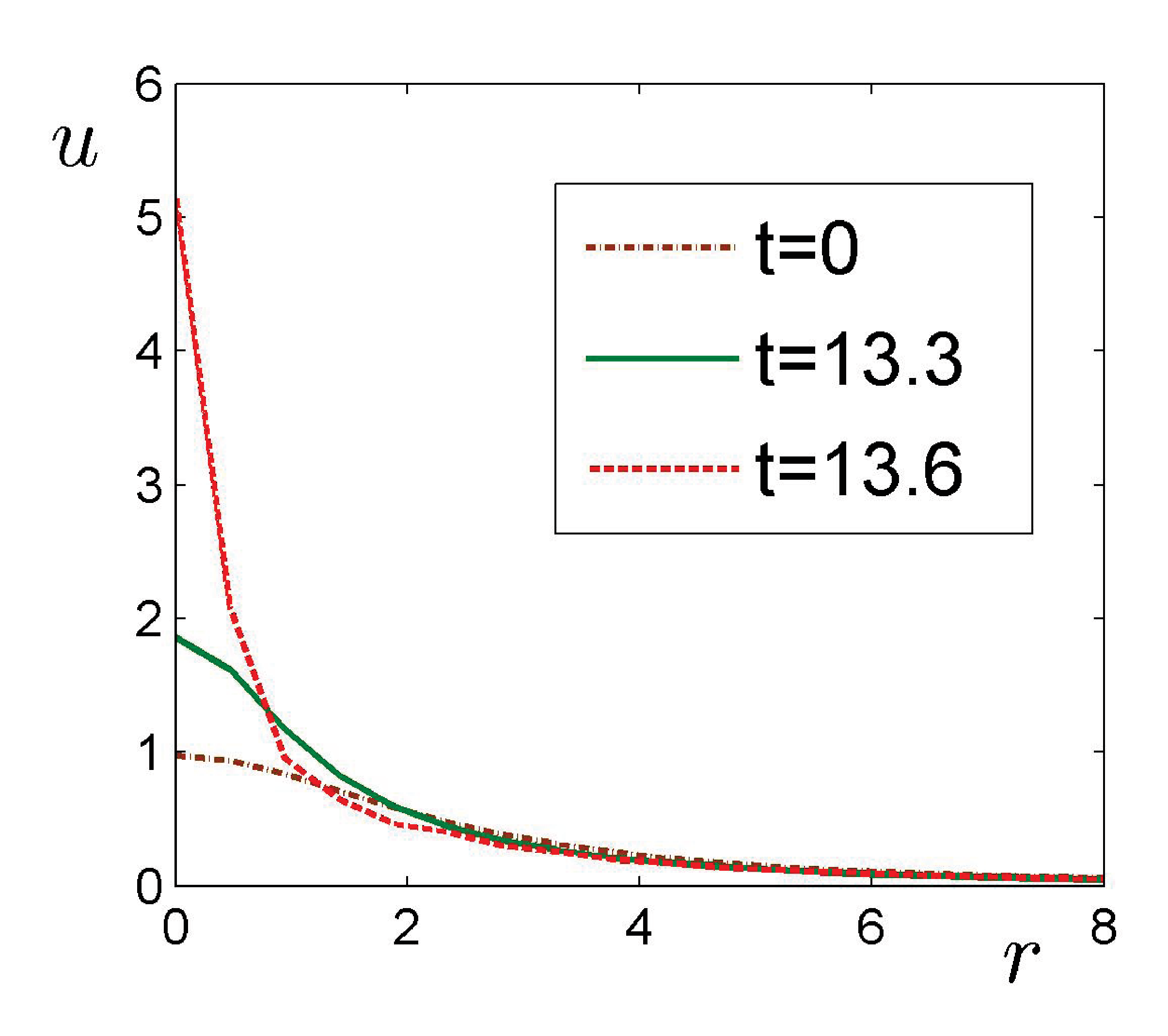}
\caption{\label{fig8} Unstable dynamics of the soliton for the
L\'{e}vy index $\alpha=1.9$ and $\lambda=0.03$; radial profiles at
the initial moment $t = 0$, at $t = 13.3$, and at $t = 13.6$.}
\end{figure}

The dependences of the 3D norm $N$ of the fundamental soliton on
the parameter $\lambda$ are shown in Fig.~\ref{fig5} for the cases
of the L\'{e}vy index $\alpha=1.2$ and $\alpha=1.9$. It can be
seen that these dependencies are nonmonotonic functions
$N(\lambda)$ of the parameter $\lambda$. The minimum points are
$\lambda_{\mathrm{cr},1}=0.057$ and
$\lambda_{\mathrm{cr},2}=0.076$ for the cases $\alpha=1.2$ and
$\alpha=1.9$, respectively. The corresponding 3D norms are
$N_{\mathrm{min},1}=84$ and $N_{\mathrm{min},2}=123$. For the case
of the L\'{e}vy index $\alpha=2$ (ordinary Laplacian), the
critical value $\lambda=0.101$  was found in \cite{Laedke1984}.
For an ordinary NLS equation ($\alpha=2$) in an arbitrary spatial
dimension $D$ and with a rather arbitrary form of nonlinearity,
the Vakhitov-Kolokolov criterion is valid. It was first formulated
in \cite{Vakhitov1973} and later rigorously justified and
generalized in \cite{Rypdal1986,Berge1998} (see also references
therein). According to this criterion, a sufficient (but not
necessary) condition for the stability of the ground state is
\begin{equation}
\label{Vakhitov} \frac{\partial N}{\partial\lambda}>0.
\end{equation}
As far as we know, the applicability of the Vakhitov-Kolokolov
criterion for a fractional NLS equation has not yet been studied.
Nevertheless, from Fig.~\ref{fig5} it is clear that for the
L\'{e}vy index $\alpha=1.2$ in the region
$\lambda>\lambda_{\mathrm{cr},1}$, as well as for the L\'{e}vy
index $\alpha=1.9$ in the region
$\lambda>\lambda_{\mathrm{cr},2}$, the corresponding ground states
satisfy the Vakhitov-Kolokolov criterion (\ref{Vakhitov}), and
therefore one may expect that they are stable.

To verify the results predicted by the Vakhitov-Kolokolov
criterion, we solved numerically the dynamical equation
(\ref{main}) initialized with our computed solutions of the form
of fundamental soliton with added Gaussian noise. The initial
condition was taken in the form $\psi
(\mathbf{x},0)=\psi_{0}(\mathbf{r})[1+\varepsilon f(\mathbf{r})]$,
where $\psi_{0}(\mathbf{r})$ is the numerically calculated
solution, $f(\mathbf{r})$ is the white Gaussian noise with
variance $\sigma^{2}=1$, and the parameter of perturbation
$\varepsilon=0.05-0.3$. The time integration was performed by the
Runge-Kutta-Merson method with the variable time step and local
error control (we used the corresponding NAG (Numerical Algorithms
Group) routine \cite{NAG}). The linear term (fractional Laplacian)
was calculated in spectral space and then transformed into
physical space. The 3D norm conserved with a relative accuracy
$<10^{-5}$ during the simulations. In the case of the L\'{e}vy
index $\alpha=1.9$, the evolution of soliton with $\lambda=0.15$,
i. e. satisfying the stability condition
$\lambda>\lambda_{\mathrm{cr},2}$, with an initial rather strong
random perturbation with $\varepsilon=0.05$ is presented in
Fig.~\ref{fig6}. It can be seen that the soliton turns out to be
robust and stable. During the evolution, the soliton undergoes
self-cleaning from the initial random disturbance. At time $t=6$,
the soliton completely restores its unperturbed shape, and then
evolves without distortion ($t=100$). Simulations with other
values of $\lambda$ in the stability region, as well as for the
L\'{e}vy index $\alpha=1.2$, also result in self-cleaning of the
soliton. If the intensity of the initial random disturbance is
sufficiently large, the soliton self-cleanses itself from noise,
simultaneously deforming, but does not undergo any significant
shape distortions. An example of such an evolution of a soliton
subjected at the initial moment of time to a strong random
perturbation with $\varepsilon=0.3$ is shown in Fig.~\ref{fig7}.
Self-cleaning of stable localized structures in the form of
multidimensional solitons and vortex solitons from initial noise
within the framework of conventional NLS equation ($\alpha=2$)
with different types of nonlinearity was observed numerically in a
number of works
\cite{Malomed2005,Malomed2006,Malomed2007PLA,Lashkin2020,Lashkin2008_78}.
Note that in our case, self-cleaning of the soliton from the
initial noise occurs for a 3D soliton with algebraically decaying
tails caused by the fractional Laplacian in the model under
consideration. The time evolution of the soliton with the L\'{e}vy
index $\alpha=1.9$ and $\lambda=0.03$, that is, corresponding to
an unstable region $\lambda < \lambda_{\mathrm{cr},2}$ in
accordance with the formal Vakhitov-Kolokolov criterion
(\ref{Vakhitov}), is shown in Fig.~\ref{fig8}. During the
evolution, the 3D norm $N$ has been conserved with an accuracy
$<10^{-7}$. Until time $ t \lesssim 13$ the soliton shape remains
unchanged. This time is equal in order of magnitude to the inverse
modulation instability growth rate $\sim 1/\lambda$ ($\lambda$
corresponds to the square of amplitude of the plane wave) of the
fastest growing mode in Eq. (\ref{max-growth}). At subsequent
moments in time $t=13.3$ and $t=13.6$, the amplitude of the
soliton rapidly increases, and at the same time soliton contracts
in accordance with the preservation of the 3D norm. That is, the
instability becomes explosive in a nonlinear regime, resulting in
a singularity in a finite time (blow-up). Thus, as predicted by
the criterion (\ref{Vakhitov}), in the unstable region the 3D
soliton collapses.

\section{Conclusion}

We have studied the fractional 3D nonlinear Schr\"{o}dinger
equation with exponential saturating nonlinearity. We have pointed
out that in the special case of the L\'{e}vy index $\alpha=1.9$,
this equation describes the dynamics of strong Langmuir turbulence
of plasma waves. The modulation instability of a plane wave has
been studied and a nonlinear dispersion equation has been derived.
In particular cases of perturbations corresponding to small and
large wavelengths compared to the length of  plane wave, the modes
of convective and absolute instability have been predicted,
respectively. The regions of instability depending on the L\'{e}vy
index, and the corresponding modulational instability growth rates
have been obtained.

Numerical solutions in the form of 3D fundamental soliton (ground
state) have been obtained for different values of the L\'{e}vy
index. The obtained dependences of the 3D norm of the soliton on
the free parameter $\lambda$ (for plasma turbulence it corresponds
to a nonlinear frequency shift) turn out to be non-monotonic and
correspond to two regions for which the Vakhitov-Kolokolov
stability criterion is formally valid. Numerical simulation has
shown that in a certain range of soliton parameters it is stable
even in the presence of a sufficiently strong initial random
disturbance, and self-cleaning of the soliton from such initial
noise has been demonstrated.

In connection with the soliton self-cleaning effect, we would like
to make a short comment. As is known, a one-dimensional NLS
equation with $\alpha=2$ and cubic nonlinearity is completely
integrable. In particular, any localized initial condition
(including a random perturbation) with a sufficiently large 1D
norm in the course of evolution results in the emergence of a pure
soliton (or several solitons) while the non-soliton part is
dispersed and carried to infinity. Therefore, in this context, the
self-cleaning of a soliton in the presence of a random disturbance
is a rigorously established fact. Multidimensional generalizations
of the NLS equation are not completely integrable, and even in
cases where the type of nonlinearity allows the existence of
stable multidimensional solitons, direct analogy is not applicable
here. Nevertheless, in our case, considering a random disturbance
against the background of a pure soliton as a localized but
dispersive wave packet, one can treat the self-cleaning as the
spreading and disappearance at infinity of the non-soliton
dispersive part in the course of evolution.

For the model under consideration, the question of the value of
the L\'{e}vy index $\alpha$ at which 3D solitons become unstable
remains open (recall that 1D solitons of the fractional NLS
equation with cubic nonlinearity collapse if $\alpha <1$). In the
case under consideration with saturating exponential nonlinearity
(as apparently for other saturating nonlinearities) these values
of the critical index depend on the parameter $\lambda$ in
contrast to the conventional model with non-saturating
nonlinearity. This problem is expected to be studied in the
future.

\section*{Declaration of competing interest}
The authors declare that they have no known competing financial
interests or personal relationships that could have appeared to
influence the work reported in this paper.

All authors have no conflicts of interest associated with this
publication, and there has been no financial support for this work
that could have influenced its outcome.

\section*{CRediT authorship contribution statement}
\textbf{V. M. Lashkin}: Conceptualization, Methodology,
Validation, Formal analysis, Investigation. \textbf{O. K.
Cheremnykh}: Conceptualization, Methodology, Validation, Formal
analysis, Investigation.

\section*{Data availability}
No data was used for the research described in the article.


\section*{References}


\begin{thebibliography}{59}

\bibitem{Oldham1974}
K. Oldham, J. Spanier, \emph{The fractional calculus} (Academic
Press, New York, 1974).

\bibitem{Samko1993}
S. G. Samko, A. A. Kilbas, and O. I. Marichev, \emph{Fractional
integrals and derivatives} (Gordon and Breach, Yverdon, 1993).

\bibitem{Miller1993}
K. S. Miller, B. Ross, \emph{An Introduction to the Fractional
Calculus and Fractional Differential Equations} (Wiley, New York,
1993).

\bibitem{Podlubny1999}
I. Podlubny, \emph{Fractional Differential Equations} (Academic
Press, San Diego, 1999).

\bibitem{Das2011}
S. Das, \emph{Functional Fractional Calculus}, 2nd edition
(Springer-Verlag, Berlin, 2011).

\bibitem{Agrawal2004}
O. P. Agrawal, J. A. Tenreiro-Machado, and I. Sabatier,
\emph{Fractional Derivatives and Their Applications: Nonlinear
Dynamics} (Springer-Verlag, Berlin, 2004).

\bibitem{Tarasov2010}
V. E. Tarasov, \emph{Fractional Dynamics: Applications of
Fractional Calculus to Dynamics of Particles, Fields and Media}
(Springer, New York, 2010).

\bibitem{Herrmann2011}
R. Herrmann, \emph{Fractional calculus: an introduction for
physicists} (World Scientific, Singapore, 2011).

\bibitem{Uchaikin2013}
V. V. Uchaikin, \emph{Fractional derivatives for physicists and
engineers} (Springer, Berlin, 2013).

\bibitem{Malomed2021}
B. A. Malomed, Optical solitons and vortices in fractional media:
a mini-review of recent results, Photonics \textbf{8}, 353 (2021).

\bibitem{Malomed2024}
B. A. Malomed, Basic fractional nonlinear-wave models and
solitons, Chaos \textbf{34}, 022102 (2024).

\bibitem{Mihalache2024}
D. Mihalache, Localized structures in optical media and
Bose-Einstein condensates: An overview of recent theoretical and
experimental results, Rom. Rep. Phys.  \textbf{76}, 402 (2024).

\bibitem{Kevrekidis2024}
\emph{Fractional Dispersive Models and Applications: Recent
Developments and Future Perspectives}, edited by P. G. Kevrekidis
and Jes\'{u}s Cuevas-Maraver (Springer, Berlin, 2024).

\bibitem{Laskin2000}
N. Laskin, Fractional quantum mechanics, Phys. Rev. E \textbf{62},
3135 (2000).

\bibitem{Laskin2002}
N. Laskin, Fractional Schr\"{o}dinger equation, Phys. Rev. E
\textbf{66}, 056108 (2002).

\bibitem{Laskin2018}
N. Laskin, \emph{Fractional Quantum Mechanics} (World Scientific
Publishing, Singapore, 2018).

\bibitem{Kwasnicki2017}
M. Kwa\'{s}nicki, Ten equivalent definitions of the fractional
Laplace operator. Fract. Calc. Appl. Anal., \textbf{20}, 7-51
(2017).

\bibitem{Oliveira2014}
E. C. de Oliveira and J. A. Tenreiro-Machado, A review of
definitions for fractional derivatives and integral, Mathematical
Problems in Engineering \textbf{2014}, 238459 (2014).

\bibitem{Davey2015}
C. Obrecht, J.-C. Saut, Remarks on the full dispersion
Davey-Stewartson equation systems, Communications on pure and
applied analysis \textbf{14}, 1547-1561 (2015).

\bibitem{Klein2014}
C. Klein, C. Sparber, and P. Markowich, Numerical study of
fractional nonlinear  Schr\"{o}dinger equations, Proc. R. Soc. A
\textbf{470}, 20140364 (2014).

\bibitem{Chen2018}
M. Chen, S. Zeng, D. Lu, W. Hu, and Q. Guo, Optical solitons,
self-focusing, and wave collapse in a space-fractional
Schr\"{o}dinger equation with a Kerr-type nonlinearity, Phys. Rev.
E \textbf{98}, 022211 (2018).

\bibitem{Zhang2017}
L. Zhang, Z. He, C. Conti, Z. Wang, Y. Hu, D. Lei, Y. Li, D. Fan,
Modulational instability in fractional nonlinear Schr\"{o}dinger
equation, Commun. Nonlinear Sci. Numer. Simulat. \textbf{48},
531-540 (2017).

\bibitem{Malomed2020}
Y. Qiu, B. A. Malomed, D. Mihalache, X. Zhu, X. Peng, Y. He,
Stabilization of single- and multi-peak solitons in the fractional
nonlinear Schr\"{o}dinger equation with a trapping potential,
Chaos, Solitons \& Fractals \textbf{140}, 110222 (2020).

\bibitem{Ovolabi2016}
K. M. Owolabi and A. Atangana, Numerical solution of
fractional-in-space nonlinear Schr\"{o}dinger equation with the
Riesz fractional derivative, Eur. Phys. J. Plus \textbf{131}, 335
(2016).

\bibitem{Ablowitz2022}
M. J. Ablowitz, J. B. Been, and L. D. Carr, Fractional integrable
nonlinear soliton equations, Phys. Rev. Lett. \textbf{128}, 184101
(2022).

\bibitem{Zakharov1972}
V.~E. Zakharov, Collapse of Langmuir waves, Sov. Phys. JETP
\textbf{35}, 908-914 (1972).

\bibitem{Thornhill1978}
S. G. Thornhill, D. ter Haar, Langmuir turbulence and modulational
instability, Phys. Rep. \textbf{43}, 43-99 (1978).

\bibitem{Goldman1984}
M.~V. Goldman, Strong turbulence of plasma waves, Rev. Mod. Phys.
\textbf{56}, 709-735 (1984).

\bibitem{Adjemyan1989}
L. Ts. Adzhemyan, A. N. Vasil'ev, M. Gnatich, and Yu. M. Pis'mak,
Quantum field renormalization group in the theory of stochastic
Langmuir turbulence, Theor. Math. Phys. \textbf{78}, 260-272
(1989).

\bibitem{Sulem1999}
C.~Sulem, P.-L. Sulem, \emph{The nonlinear Schr\"{o}dinger
equation: self-focusing and wave collapse} (Springer-Verlag, New
York, 1999).

\bibitem{Laedke1984}
E. W. Laedke and K. H. Spatschek, Stable three-dimensional
envelope solitons, Phys. Rev. Lett. \textbf{52}, 279 (1984).

\bibitem{Lashkin2020}
V. M. Lashkin, Stable three-dimensional Langmuir vortex soliton,
Phys. Plasmas \textbf{27}, 042106 (2020).

\bibitem{Kivshar_book2003}
Y.~S. Kivshar and G.~P. Agrawal, \emph{Optical Solitons: From
Fibers to Photonic Crystals} (Academic Press, San Diego, 2003).

\bibitem{Qiu2020}
Y. Qiu, B. A.  Malomed, D. Mihalache, X. Zhu, L. Zhang, Y. He,
Soliton dynamics in a fractional complex Ginzburg-Landau model,
Chaos Solitons Fractals \textbf{131}, 109471 (2020).

\bibitem{Frank2013}
R. L. Frank, E. Lenzmann, Uniqueness of nonlinear ground states
for fractional Laplacians in $\mathbb{R}$, Acta Math.
\textbf{210}, 261-318 (2013).

\bibitem{Frank2016}
R. L. Frank, E. Lenzmann, and L. Silvestre, Uniqueness of radial
solutions for the fractional Laplacian, Comm. Pure Appl. Math.
\textbf{69}, 1671-1726 (2016).

\bibitem{Li2024}
X. Lia and L. Song, Uniqueness of positive solutions for
fractional Schr\"{o}dinger equations with general nonlinearities,
arXiv:2401.02795 [math.AP].

\bibitem{Petviashvili1976}
V. I. Petviashvili, Equation of an extraordinary soliton, Sov. J.
Plasma Phys. \textbf{2}, 257-258 (1976).

\bibitem{Petviashvili_book1992}
O.~A. Pokhotelov and V.~I. Petviashvili, \emph{Solitary Waves in
Plasmas and in the Atmosphere}  (Gordon and Breach, Reading,
1992).

\bibitem{Lakoba2007}
T. I. Lakoba, J. Yang, A generalized Petviashvili iteration method
for scalar and vector Hamiltonian equations with arbitrary form of
nonlinearity, J. Comput. Phys. \textbf{226}, 1668-1692 (2007).

\bibitem{Lashkin2008_77}
V. M. Lashkin, Two-dimensional multisolitons and azimuthons in
Bose-Einstein condensates, Phys. Rev. A \textbf{77}, 025602
(2008).

\bibitem{Lashkin2008_78}
V. M. Lashkin, Stable three-dimensional spatially modulated vortex
solitons in Bose-Einstein condensates, Phys. Rev. A \textbf{78},
033603 (2008).

\bibitem{Pelinovsky2004}
D. E. Pelinovsky, Yu. A. Stepanyants, Convergence of
Petviashvili's iteration method for numerical approximation of
stationary solutions of nonlinear wave equations, SIAM J. Numer.
Anal. \textbf{42} 1110-1127 (2004).

\bibitem{Vakhitov1973}
M. G. Vakhitov and A. A. Kolokolov, Stationary solutions of the
wave equation in the medium with nonlinearity saturation,
Radiophys. Quantum Electron. \textbf{16}, 783 (1973).

\bibitem{Rypdal1986}
J. J. Rasmussen and K. Rypdal, Blow-up in nonlinear Schroedinger
equations-I: A general review, Physica Scripta \textbf{33},
481-497 (1986).

\bibitem{Berge1998}
L. Berg\'{e}, Wave collapse in physics: principles and
applications to light and plasma waves, Phys. Rep. \textbf{303},
259-370 (1998).

\bibitem{NAG}
\emph{NAG Fortran Library, Mark 18} (Numerical Algorithms Group,
Oxford, 1999).

\bibitem{Malomed2005}
B. A. Malomed, D. Mihalache, F. Wise, and L. Torner,
Spatiotemporal optical solitons, J. Opt. B: Quantum Semiclass.
Opt. \textbf{7}, R53-R72 (2005).

\bibitem{Malomed2006}
D. Mihalache, D. Mazilu, F. Lederer, Y. V. Kartashov, L.-C.
Crasovan, L. Torner, and B. A. Malomed, Stable vortex tori in the
three-dimensional cubic-quintic Ginzburg-Landau equation, Phys.
Rev. Lett. \textbf{97},  073904 (2006).

\bibitem{Malomed2007PLA}
B.~A. Malomed, F. Lederer, D. Mazilu, D. Mihalache, On stability
of vortices in three-dimensional self-attractive Bose-Einstein
condensates, Phys. Lett. A \textbf{361}, 336-340 (2007).


\end{thebibliography}
\end{document}